\documentclass[
12pt,
3p,
review,
]{elsarticle}
\usepackage{hyperref}
\usepackage{graphicx,graphics}
\usepackage{placeins}
\usepackage{amsmath}
\usepackage{amssymb}
\usepackage{mathtools}
\usepackage{array}
\usepackage{hyperref}
\usepackage{color}
\usepackage{soul}
\usepackage{dcolumn}
\usepackage{bm}
\usepackage{todonotes}
\usepackage[utf8]{inputenc}
\usepackage[english]{babel}
\usepackage{libertine}
\usepackage{pgfplots}
\usepgfplotslibrary{polar}
\pgfplotsset{compat=1.9}
\usetikzlibrary{calc}

\usepackage[title]{appendix}

\usepackage{multirow}

\usepackage{layouts}

\usepackage{caption}
\usepackage{subcaption}

\newcommand{\vphi}{\mbox{{\bm{$\phi$}}}}

\newcommand{\vc}{\mbox{{\bm{$c$}}}}

\newcommand{\vv}[1]{\boldsymbol{#1}}
\newcommand{\n}{\vv{\nabla}}
\renewcommand{\l}{\mathopen{}\mathclose\bgroup\left}
\renewcommand{\r}{\aftergroup\egroup\right}

\newcommand{\diff} [1]{\mathrm{d}{#1}} 
\newcommand{\phia}{\phi_{\alpha}}
\newcommand{\phib}{\phi_{\beta}}

\let\originaleps=\epsilon
\let\epsilon=\varepsilon
\let\varepsilon=\originaleps
\usepackage{stmaryrd}

\mathchardef\hy="2D

\definecolor{mydark_blue}{RGB}{0, 0, 139}
\definecolor{myblue}{RGB}{0, 0, 255}
\definecolor{mycyan}{RGB}{0, 255, 255}  
\definecolor{mygreen}{RGB}{0, 255, 0}
\definecolor{myyellow}{RGB}{255, 255, 0}
\definecolor{myred}{RGB}{255, 0, 0}
\definecolor{mydark_red}{RGB}{139, 0, 0}
\definecolor{myblack}{RGB}{0, 0, 0}

\definecolor{BRY_1}{RGB}{  0,  0,255}
\definecolor{BRY_2}{RGB}{127,  0,127}
\definecolor{BRY_3}{RGB}{255,  0,  0}
\definecolor{BRY_4}{RGB}{255,127,  0}
\definecolor{BRY_5}{RGB}{255,255, 85}

\begin{document}

\begin{frontmatter}
\title{An alternate approach for estimating grain-growth kinetics}

\author[mymainaddress]{M Prabhakar}
\author[mymainaddress,mysecondaryaddress]{P G Kubendran Amos\corref{mycorrespondingauthor}}
\ead{prince@nitt.edu}

\cortext[mycorrespondingauthor]{P G Kubendran Amos}

\address[mymainaddress]{Theoretical Metallurgy Group,
Department of Metallurgical and Materials Engineering,\\
National Institute of Technology Tiruchirappalli, \\
Tamil Nadu, India}

\address[mysecondaryaddress]{Institute of Applied Materials (IAM-MMS),
Karlsruhe Institute of Technology (KIT),\\
Strasse am Forum 7, 76131 Karlsruhe, Germany
}

\begin{abstract} 


\textcolor{black}{
Rate of grain growth, which aides in achieving desired properties in polycrystalline materials, is conventionally estimated by measuring the size of grains and tracking its change in micrographs reflecting the temporal evolution. 
Techniques adopting this conventional approach demand an absolute distinction between the grains and the interface separating them to yield an accurate result. 
Edge-detection, segmentation  and other deep-learning algorithms are increasingly adopted to expose the network of boundaries and the associated  grains precisely. 
An alternate approach for measuring grain-growth kinetics, that curtails the need for advanced image-processing treatment, is presented in this work. 
Grain-growth rate in the current technique is ascertained by \textit{counting} the number of triple-( and quadruple-) junctions, and monitoring its change during the microstructural evolution. 
The shifted focus of this junction-based treatment minimises the significance of a well-defined grain-boundary network, and consequently, the involvement of the sophisticated techniques that expose them. 
A regression-based object-detection algorithm is extended to realise, and count, the number of junctions in polycrystalline microstructures. 
By examining the change in the number of junctions with time, the growth rate is subsequently determined.
Growth kinetics estimated by the present junction-based approach, across a wide-range of multiphase polycrystalline microstructures, agree convincingly with the outcomes of the conventional treatment.
Besides offering a novel technique for grain-growth measurement, the analysis accompanying the current work unravels a trend, compatible with the topological events, in the progressive evolution of the triple-junctions count.
The present approach, through its underlying algorithm, provides a promising option for monitoring grain-growth during in-situ investigations.
}

\end{abstract}


\end{frontmatter}


\section{Introduction}

The unique, and often, striking properties of a nanocrystalline material are primarily the outcome of the characteristic size of its constituent \lq crystals \rq \thinspace or grains.
In an application, wherein a nanocrystalline material is employed, the span of its applicability is essentially the duration across which the characteristic grain-size is retained~\cite{koch2008stabilization}. 
In other words, under the associated conditions, rate of grain growth determines the life of a nanocrystalline component in any application. 
Sluggish the grain growth, increased is the life span of nanocrystalline material~\cite{pande2005grain}. 
This aspect of nanomaterials becomes extremely critical in high temperature applications~\cite{koch2013high}. 
Besides these specific category of materials, properties of polycrystalline systems, ranging from alloys to ceramics, are noticeably influenced by grain size~\cite{weissmuller1993alloy,kingery1963review}.  
Correspondingly, processing techniques are developed to establish the desired grain size, ultimately yielding the required properties~\cite{sahay2003accelerated}. 
Designing of such processing techniques demands a quantitative understanding of the grain-growth kinetics.
Owing to these, and several other, critical reasons different approaches have been developed and adopted to measure the rate of grain growth. 

The techniques employed to ascertain the rate of grain growth can broadly be categorised into two. 
In the first approach, the kinetics are estimated by measuring the average size of grains at specific instances during the temporal evolution of polycrystalline microstructures. 
The techniques associated with this approach can be referred to as \lq static \rq \thinspace, since the growth rate is determined from the temporally-fixed microstructures (images) captured at different instances. 
Two well-known example of the static approaches are line-intercept~\cite{han1998determination} and planimetric methods~\cite{khzouz2011grain}. 
A line of specific length is introduced in a polycrystalline micrograph, in the line-intercept method, and the number of intersections between the grain boundaries and the line is counted. 
In essence, finer the grains, larger the number of intersections. 
By taking the reciprocal of the number of intersections, the intercept length is calculated. 
A statistically more rigorous measure, called average intercept length ($l(t)$), is ascertained from several parallel lines which is subsequently translated to ASTM grain-size number~\cite{subcommittee1996standard}.
While following the similar approach of gathering relevant information from static representation of microstructure, planimetric method estimates the grain size by focusing on the number of complete grains encapsulated in test section of a definite area~\cite{astm2015standard}. 
The average grains per unit area ($N_A$) is estimated from several sections of identical area across the polycrystalline micrograph. 
Through the relation, $G(t) = 3.322[\text{log}~N_A(t)] - 2.954,$ the average grain size, G(t), is calculated.
Measuring the average grain size of different polycrystalline micrographs, captured across the temporal evolution, ultimately yields the growth kinetics in the static approach. 

Owing to the straightforward framework, the static approaches are widely adopted for measuring the average grain size in polycrystalline systems. 
However, these techniques are rather arduous and demand scrupulous labor. 
Furthermore, the accuracy of the outcome rendered by the static techniques depends solely on the efficacy of etching. 
When the grain boundaries are not sufficiently distinguished, the size measurement becomes rather compromised. 
Besides, given its underlying framework, the static techniques cannot be adopted to measure growth kinetics during in-situ observation. 
These limitations of static methods have led to the development of the second category of \lq dynamic \rq \thinspace techniques to ascertain grain growth kinetics.

The dynamic approach characteristically measures the grain-growth kinetics directly, as opposed to ascertaining it from the static representation of micrographs captured at progressive instances.
Correspondingly, the dynamic techniques are fundamentally different and more sophisticated. 
One such sophisticated dynamic method involves ultrasonic waves to measure the rate of grain growth~\cite{maalekian2012situ}. 
A material exposed to ultrasonic waves largely allows the waves to pass through unimpeded.
Any deviation from a propagation pathway is observed only when the waves interact with an inhomogeneity. 
Polycrystalline systems are fraught with grain boundaries that introduce inhomogeneities by interrupting the specific alignment of atoms. 
Therefore, the scattering of ultrasonic waves can be quantitatively related to the area of grain boundaries in a material. 
Stated otherwise, the amount of ultrasonic waves scattered by a fine-grained material is noticeably greater than a coarse-grained system. 
Consequently, as the grains grow in a polycrystalline system, the scattering of ultrasonic waves gradually reduces, and propagation becomes largely unperturbed. 
Exploiting this mechanism, the growth kinetics of grains, in the corresponding dynamic technique, is estimated by monitoring the rate of progressive decrease in the scattering of the ultrasonic waves.
Although this method renders a real-time estimation of the grain-growth rate, it demands an appropriate source for the invasive ultrasonic waves. 
The combined involvement of the pulsed laser and the interferometer, which serve as the source and detector of the ultrasonic waves, contributes to the intricacy of the technique, thereby ultimately limiting its applicability. 

In another dynamic approach, the change in enthalpy accompanying the grain growth is measured and subsequently translated to the kinetics~\cite{chen1991analysis}. 
Grain growth in the polycrystalline system is characterized by a progressive decrease in the density of the grain boundaries. 
The increasing disappearance of the sections of grain boundaries induces an exothermic change in the interfacial enthalpy. 
This enthalpy change can be measured and tracked by differential scanning calorimetry~\cite{zhilyaev2002calorimetric}. 
By monitoring the calorimetric signal, which reflects the rate of enthalpy change, the kinetics of grain growth is determined in this dynamic technique. 
The underlying principle of this technique allows the measurement of growth kinetics across a given material, in its entirety, as opposed to specific sections. 
However, this approach has a telling limitation in its usage. 
The enthalpy change accompanying the disappearance of interfaces is generally imperceptible and cannot be easily detected. 
Consequently, the enthalpy-based method becomes viable only when the change in the grain-boundary density is significant, as typically observed in nanocrystalline materials.

\textcolor{black}{Advancements made in artificial intelligence are increasingly employed to understand the behavior of materials~\cite{sha2020artificial, holm2020overview,ni2023physics}. }
Some of these advancements have recently been employed to aid in measuring grain-growth kinetics.
Considering that the inadequate realization of grain boundaries is a principal source of inaccuracy in estimating grain size, and thereby growth rate, deep learning techniques have been employed to aid the estimation of kinetics.
In a recent work, an edge detection algorithm is adopted to distinguish the existing grain boundaries in a polycrystalline microstructure~\cite{banerjee2019automated}. 
Upon distinguishing the grain boundaries, the associated grains are realised as closed contours by removing any discontinuities in the interface network and removing any spurious edges. 
The static measurement techniques, in an automated framework, are extended to the deep-learning-enhanced polycrystalline micrographs for accurate measurement of the grain size. 
Besides this approach, a technique estimating the grain size from the intersection of the boundaries in the \lq topological-skeleton\rq \thinspace has subsequently been reported~\cite{li2022automation}.
Furthermore, a combination of object detection and segmentation has been employed in a later study to estimate the grain size~\cite{gorynski2023machine}. 
These recent advancements essentially reinforce the preceding works wherein an absolute distinction between the boundary, grains and any additional phases is established largely by 
\begin{enumerate}
 \item introducing a characteristic threshold for every microstructure to separate the grains from the rest~\cite{peregrina2013automatic}
 \item employing a combination of segmentation with appropriate pre- and post-processing techniques~\cite{campbell2018new}
 \item a combination of segmentation and adaptive thresholding \cite{sosa2014development}.
\end{enumerate}

\textcolor{black}{Highlights of the existing works are listed in Table ~\ref{tab:list}.}
\begin{table*}
  \caption{\textcolor{black}{Chronological summary of the reported approaches} \label{tab:list}.} 
 \centering
 \resizebox{\columnwidth}{!}
 {
 \begin{tabular}{c | c | c }
  Work 		& Technique 	& Grain size measurement \\ [0.5ex]
  \hline
  H. Peregrina-Barreto et al. (2013)~\cite{peregrina2013automatic}	&  Automatic Digital Sample Preparation (image processing) 				& Planimetric count  \\
IA. Campbell et al. (2018)~\cite{campbell2018new} 	& Watershed transform with pre- and post-processing	&  Aspect-ratio based estimation \\
S. Banerjee et al. (2019)~\cite{banerjee2019automated} & Edge detection and closed contour formation & Intercept and Planimetric method \\
X. Li, L. Cui, J. Li et al. (2022)~\cite{li2022automation} &    Topological-skeleton based algorithm    &  Intercept method \\
C. Gorynski et al. (2023)~\cite{gorynski2023machine} & Mask R-CNN & Pixel-based (Mask) calculation\\
  \hline
 \end{tabular}
 }
\end{table*} 
Irrespective of the differences in the microstructure  processing, growth kinetics in all the existing works are ascertained by first estimating the accurate grain size through an established \lq static \rq \thinspace technique and subsequently tracking its evolution.
Such framework for estimating growth rate, though renders an accurate outcome which can be related to existing standards, largely overlooks the evolution of other microstructural features that accompany grain growth. 
Therefore, deviating from the principal focus on grain size, and employing a suitable artificial intelligence (AI) tool, the present work seemingly offers a novel approach for measuring grain-growth kinetics. 

Any polycrystalline microstructure encompasses following two characteristic features:
\begin{enumerate}
 \item grain boundaries, the interface separating the grains
 \item largely zero-dimensional points called triple-junctions, wherein grain boundaries meet. 
\end{enumerate}
Quadruple-junctions are another feature that accompanies grain boundaries and triple-junctions in extremely anisotropic polycrystalline systems~\cite{mora2008effect}. 
While techniques to ascertain growth kinetics predominantly focus on the evolution of grain boundaries, in the present work, a triple-junction-based attempt is made to comprehend the rate. Accordingly, the current work comprises two sections. 
In the first section, an object-detection tool is developed to identify and count both triple- and quadruple-junctions in a polycrystalline microstructure. 
This tool is subsequently used in the second section to delineate the grain-growth kinetics.

\section{Detecting Triple- and Quadruple-Junctions}

\subsection{Overview}

\begin{figure}
    \centering
      \begin{tabular}{@{}c@{}}
      \includegraphics[width=0.9\textwidth]{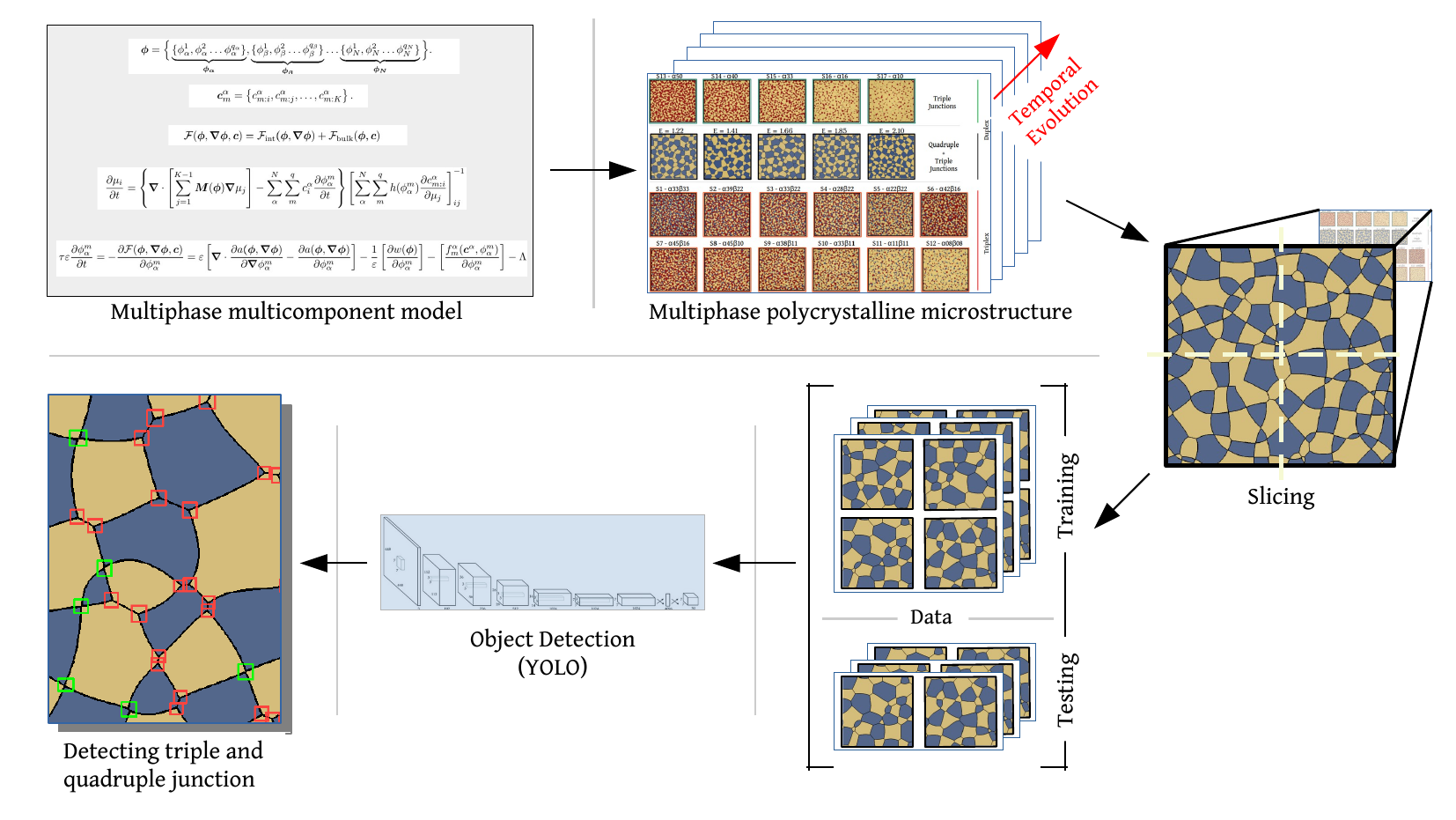}
    \end{tabular}
    \caption{ A graphical depiction of the approach adopted to develop a regression-based tool for detecting quadruple- and triple-junctions. Multiphase polycrystalline micrographs exhibiting grain growth rendered by multiphase-field model is involved in training and validating the tool for realising triple- and quadruple-junctions.
    \label{fig:technique}}
\end{figure}

The approach employed in the present work to devise an AI tool for recognising and counting triple- and quadruple-junctions in polycrystalline systems is illustrated in \ref{fig:technique}.
In other words, triple- and quadruple-junctions of the polycrystalline microstructures are designated as unique \textit{objects}, and are subsequently realised by an appropriately equipped object-detection tool. 
A known object-detection algorithm referred to as You-Only-Look-Once (YOLOv5) plays a central role in realising the desired features of the polycrystalline microstructure. 

As opposed to other object-detection algorithms, YOLO operates on a regression-based framework~\cite{jiang2022review}. 
Simply stated, for a given polycrystalline microstructure, a prediction is made by YOLO on the type and position of the junction. 
The accuracy of this stochastic prediction is gradually enhanced, across several iterations, by minimising the deviation from the labelled data~\cite{du2018understanding}. 
Repeating this process of prediction and its subsequent refinement over different sets of microstructures essentially \textit{trains} the algorithm to accurately realise the junctions in unforeseen micrographs. 
The efficiency of this training, which dictates the performance of the tool, depends on the size of the dataset solely available for training. 
This need for a corpus of data is catered, in the present work, by wide-range of numerically generated polycrystalline microstructures. 
A tessellation scheme, which has been proven to introduce physically-relevant configuration of grains, is adopted to develop the polycrystalline microstructures~\cite{suzudo2009evolutional}. 
These microstructures are allowed to evolve, within the framework of multicomponent multiphase-field model~\cite{amos2020multiphase}, thereby rendering sequence of micrographs exhibiting grain growth and further increasing the size of the dataset. 
(The multicomponent multiphase-field adopted in the present work is briefly discussed in Appendix 1~\ref{App:1}.) 
Correspondingly, the dataset includes $100$ micrographs, for every original numerically-generated polycrystalline microstructure, capturing the different instances of the simulated grain growth. 
For the ease of handling, both original and evolved microstructures are sliced into four sections as shown in \ref{fig:technique}. 
Since microstructures are devised to include atleast few hundred grains at any given instant, even the sliced micrographs possess sufficient information for training the algorithm. 
The dataset encompassing sliced original- and evolved-microstructures is split into training and validation set in the proportion 80\% and 20\%, respectively. 
The object detection algorithm is trained to identify and count the triple- and quadruple-junctions from the predetermined larger portion of the dataset, while validation set is exploited to fine-tune the accuracy through the hyperparameters. 

\subsection{Principal Dataset}

\begin{figure}
    \centering
      \begin{tabular}{@{}c@{}}
      \includegraphics[width=1.0\textwidth]{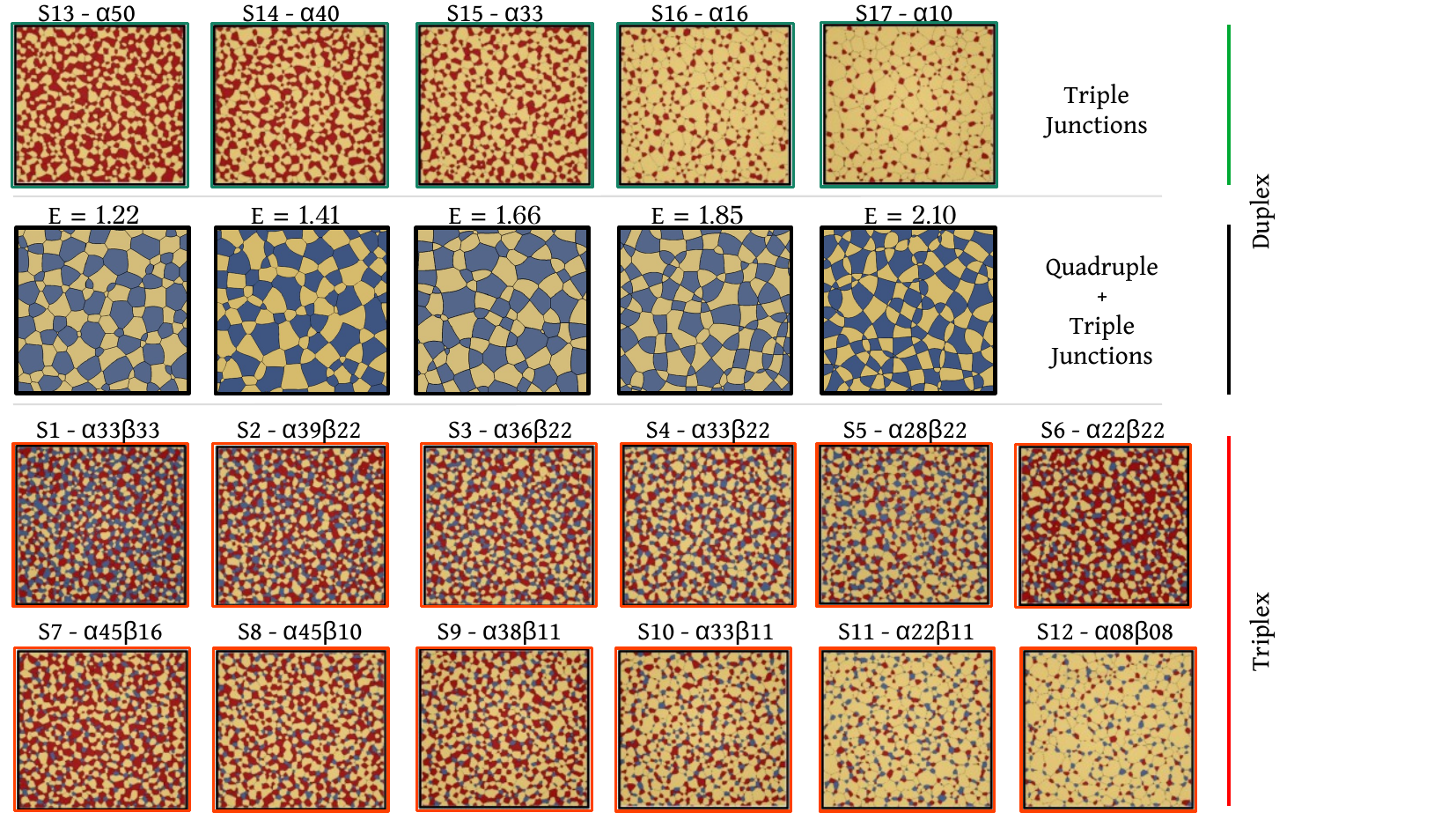}
    \end{tabular}
    \caption{ The principal dataset for the present investigation includes multiphase polycrystalline microstructures of various phase-fractions. The numerically-generated microstructures are turned to multiphase systems with varying phase-fractions, and are allowed to evolve, within the framework of multiphase-field model. The fundamental polycrystalline microstructure, along with its temporal evolution, constitute the dataset for the present investigation. These microstructures are designated primarily based on the volume-fraction of the minor-phases. A set of anisotropic duplex-microstructures are devised to introduce stable quadruple junctions.
    \label{fig:microstructures}}
\end{figure}

A statistically rigorous proof of the present junction-based approach in ascertaining growth rate is achieved by testing it against microstructures exhibiting varied kinetics. 
Correspondingly, a dataset is devised which primarily encompasses varied microstructures rendering characteristic grain-growth kinetics. 
In these microstructures, a perceivable difference in the growth kinetics is achieved by considering multiphase systems, as opposed to homogeneous single-phase microstructure.
Grains constituting the microstructures are associated with one of two or three phases thereby establishing polycrystalline duplex or triplex system.
A characteristic phase-fraction is introduced in a microstructure by altering the number of grains pertaining with each phase. 

Fig.~\ref{fig:microstructures} shows all duplex and triplex systems with  unique and noticeably distinct phase-fractions considered in the present study.
Since the noticeable difference in the grain growth kinetics stems primarily from the varying phase-fractions, isotropic grain-boundary energy is assumed across the various duplex and triplex microstructures. 
In other words, for most polycrystalline microstructures in Fig.~\ref{fig:microstructures}, identical and representative grain boundary energy is assigned to the interfaces separating grains of both similar and dissimilar phases. 
Interfacial energy and other parameters characterising the duplex and triplex microstructures are tabulated in Tables~\ref{tab:mat1},~\ref{tab:mat2} and~\ref{tab:mat3} included Appendix 1~\ref{App:1}. 

Under isotropic energy conditions,  triple junctions are the only other stable feature in polycrystalline microstructures, besides grain boundaries. 
Though quadruple junctions may form during grain growth as a result of the disappearance of grains, these features eventually dissociate to energetically-stable triple junctions. 
In order to explicitly introduce stable quadruple junctions, specific anisotropies in interfacial energies are introduced in duplex systems with equifraction of phases, as shown in Fig.~\ref{fig:microstructures}. 
In these two-phase microstructures, anisotropy is achieved by assigning different grain-boundary energies to the interface separating grains of like phases, when compared to the dissimilar grains~\cite{perumal2020quadrijunctions}.
Put simply, while the representative value of the energy is left unaltered for the interface between the dissimilar grains, a specific value is associated with the boundaries of same-phase grains. 
Based on these two types of interfacial energies, the overall anisotropy of the microstructure is quantified as $E_\alpha=\frac{\gamma_{\alpha\alpha}}{\gamma_{\alpha\beta}}$ and  $E_\beta=\frac{\gamma_{\beta\beta}}{\gamma_{\alpha\beta}}$. 
Given the focus to introduce stable quadruple-junctions, in the present study $E_\alpha=E_\beta(=E)$ is assumed. 

As shown in Fig.~\ref{fig:microstructures}, five different anisotropic conditions are considered by varying energies of interface separating similar grains. 
As the value of grain-boundary-energy ratio (E) increases, the number of quadruple junctions proportionately increases~\cite{perumal2020quadrijunctions}. 
Similar to other duplex and triplex microstructures, these anisotropic systems are allowed to evolve, within the framework of multicomponent multiphase-field model, solely to increase the wealth of data with stable quadruple junctions. 

\subsection{Training and Validation}

\begin{figure}
    \centering
      \begin{tabular}{@{}c@{}}
      \includegraphics[width=1.0\textwidth]{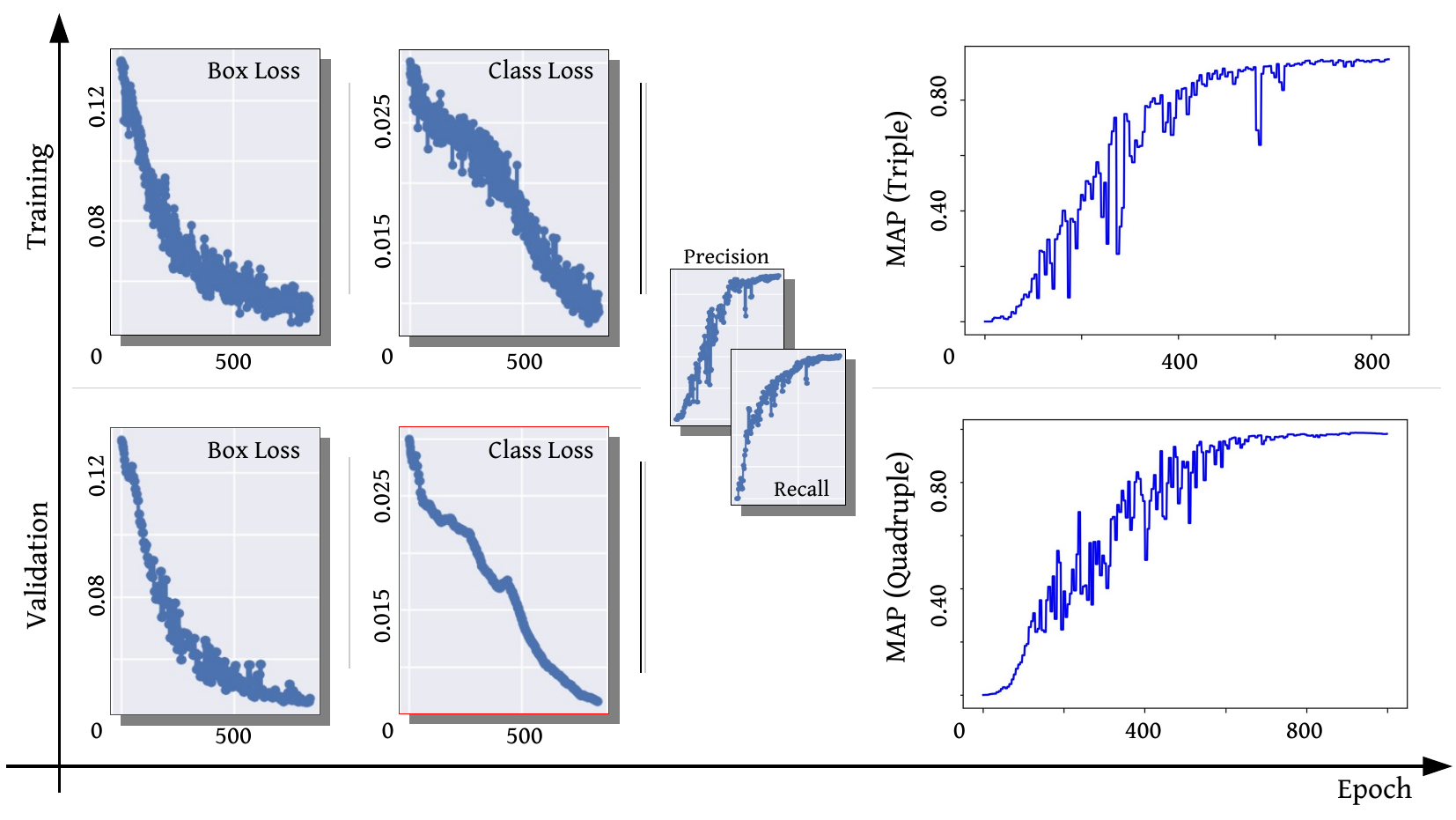}
    \end{tabular}
    \caption{ Training and validation of the object-detection technique in identifying triple- and quadruple junctions as different class of objects in polycrystalline microstructures. The progressive improvement in the performance of the algorithm in the present framework is assessed through loss functions and mean average precision (MAP).
    \label{fig:metrics}}
\end{figure}

The tool to identify and count triple- and quadruple-junctions in polycrystalline microstructures is developed by training and validation of the object-detection algorithm.
The translation of a generic detection-algorithm to a junctions-realising tool, during training and validation, is grasped through suitable metrics and illustrated in Fig.~\ref{fig:metrics}.
This illustration also includes a parameter indicating the overall performance of the junction-detection tool on unforeseen microstructures. 
Epoch in Fig.~\ref{fig:metrics} is related to the iterations through batch size, which is considered to be $4$ in the present work. 

Detecting a junction in a polycrystalline microstructure begins with a characteristic prediction made by the underlying algorithm. 
This prediction assumes a form of the \textit{bounding box} which supposedly encompassed the junction. 
The predicted bounding box is encoded as a vector with entries representing the coordinates of its position, dimension and probability of locating a junction within its encapsulation. 
The mismatch between the prediction and ground truth, gathered through manual labelling, is quantified as localization- and confidence-loss. 
While the localization loss describes the deviation in features of the predicted bounding-box and manual label, the probability of an anticipated detection in comparison to the certainty of the ground truth is expressed as confidence loss. 
The localization and confidence loss cumulatively renders the \textit{box loss} illustrated in Fig.~\ref{fig:metrics}. 
Exposing the algorithm to sufficient number of polycrystalline microstructures with manually-labelled triple and quadruple junctions, over several epochs, trains it to detect these microstructural features.
Efficient training of the algorithm is indicated by the noticeable decrease in the box loss with epochs. The performance of this trained junction-detecting tool on unforeseen and relatively smaller dataset is analysed during validation. 
After sufficient epochs, the accuracy of the approach in recognising triple- and quadruple-junctions across the unknown polycrystalline structures noticeably improves. 

While box loss focuses on accurately locating and encasing the junction, the distinction between the types of junction is achieved through the \textit{class loss}. 
The confidence score which reflected the probability of locating any junction within the predicted box is extended, and individual scores are included to account for triple- and quadruple-junctions separately.
These separate confidence scores indicate the probability of finding a triple- or quadruple-junction with a bounding box. 
In other words, the algorithm through different confidence scores are trained to distinguish triple junction from the quadruple points. 
Any inaccuracy in detecting the type of the junctions, in view of the ground truth, is quantified as class loss. 
Similar to the box loss, the mismatch in prediction and ground truth pertaining to the type of junctions is weeded out through sufficient training across several epochs. 
The ability to distinguish triple from quadruple junction in previously unknown polycrystalline microstructure increases with epochs, as indicated by the decrease in class loss in Fig.~\ref{fig:metrics}. 
During the training of the algorithm, the hyperparameters are optimised. 
Correspondingly, the values 0.01, 0.0005 and 0.937 are assigned to learning rate, weight decay and momentum, respectively~\cite{prabakar2023regression,venkatanarayanan2023accessing}.

\begin{figure}
    \centering
      \begin{tabular}{@{}c@{}}
      \includegraphics[width=1.0\textwidth]{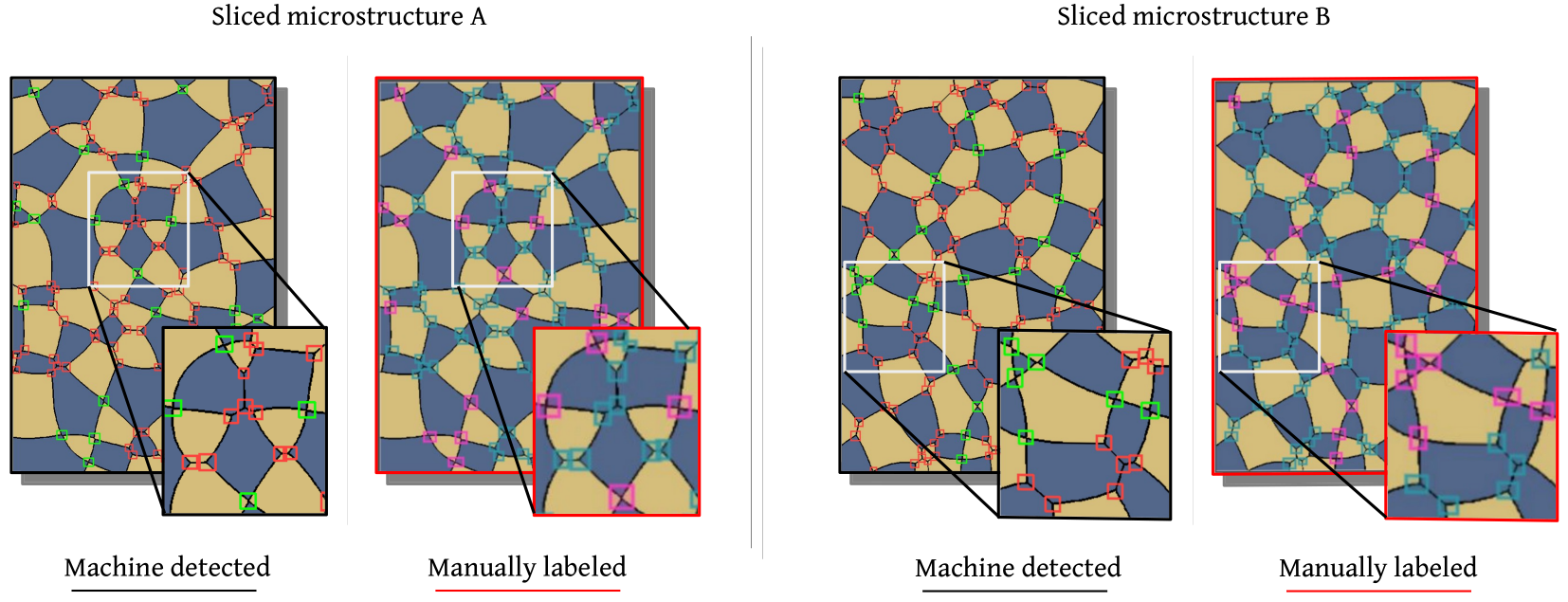}
    \end{tabular}
    \caption{ A direct side-by-side comparison of the ground-truth and detection of the trained algorithm on duplex microstructures. The ground truth on the right of each window is established through manual labelling. Triple- and quadruple-junctions of the anisotropic duplex microstructures are distinctly labelled, and this distinction is accurately captured by the trained technique. 
    \label{fig:visual}}
\end{figure}

In addition to comprehending training and validation, the overall performance of the junction-realising tool over unforeseen microstructures is studied through suitable parameters. 
The fundamental parameters adopted to unravel the overall accuracy of the  junction-tool are \textit{Recall} and \textit{Precision}.
By considering the ratio of true positives to the sum of true and false positives, precision relates the accurate detection of the junctions to the inaccurate realisation.
Recall, on the other hand, indicates the accurate recognition in view of the failed detection by normalising true positives with the some of true positives and false negatives. 
Considering that the formulation of recall includes failed detection, and precision encompasses incorrect realisation, the optimisation of these parameters, enhances the overall performance of the technique.
This optimisation, which improves the accuracy of the detections, is expressed as a precision-recall curve.  
The trajectory of the precision-recall curve varies with the class of objects. 
Correspondingly, the performance of the junction-tool is comprehensively gathered by two precision-recall curves separately indicating the accurate identification of triple- and quadruple-junctions.
The area encapsulated by the precision-recall curve, for a given junction (object), is referred to as the mean average precision (MAP). 
The improvement in mean average precision of the tool in realising triple- and quadruple-junctions is separately shown in Fig.~\ref{fig:metrics}.
Given the description of mean average precision, MAP=1.0 is the maximum value it can assume. 
An absolute MAP (=1.0) value indicates a perfect performance of the junction-realising tool characterised no inaccurate or missed detection. 
The present approach through the optimisation of precision and recall offers mean average precision of 0.897 and 0.947 for triple- and quadruple-junctions, respectively. 
Considering the proximity of the observed value to the perfect score, the performance of the developed technique is deemed convincing and is adopted for analysing the microstructural changes. 

A more straightforward representation of the performance exhibited by the present approach in identifying junctions in polycrystalline microstructures is offered in Fig.~\ref{fig:visual}. 
This illustration compares the manual labelling, which establishes the ground truth, to the detections of the developed tool. 
Duplex system with grain-boundary anisotropy is specifically chosen for this representation, given the presence of both triple- and quadruple-junctions in its microstructure. 
It is evident from Fig.~\ref{fig:visual} that the algorithm is adequately trained, and is capable of distinguishing  triple- and quadruple-junctions in a microstructure, besides collectively recognising them in an intricate polycrystalline microstructure.

\section{Ascertaining grain-growth kinetics}

\subsection{Inverse triple-junction plots}

The junction-detection tool developed in the previous section is adopted to analyse series of microstructures indicating temporal evolution of grains in multiphase polycrystalline systems. 
Grain growth exhibited by all duplex and triple microstructures included in Fig.~\ref{fig:microstructures}, except anisotropic systems with stable quadruple-junctions, are considered for this investigation. 
Though the tool is capable of handling the quadruple junctions, systems with grain-boundary anisotropy are rather overlooked in the present analysis for two reasons.
One, considering varied phase-fractions in isotropic system, a noticeable disparity in growth kinetics is inherently expected. 
Secondly, a comparison of grain-growth rate across the different microstructures is manageable and consistent when the anisotropic systems are excluded. 
Moreover, observing measurable amount of stable quadruple junctions in physical system is extremely rare. 

\begin{figure}
    \centering
      \begin{tabular}{@{}c@{}}
      \includegraphics[width=1.0\textwidth]{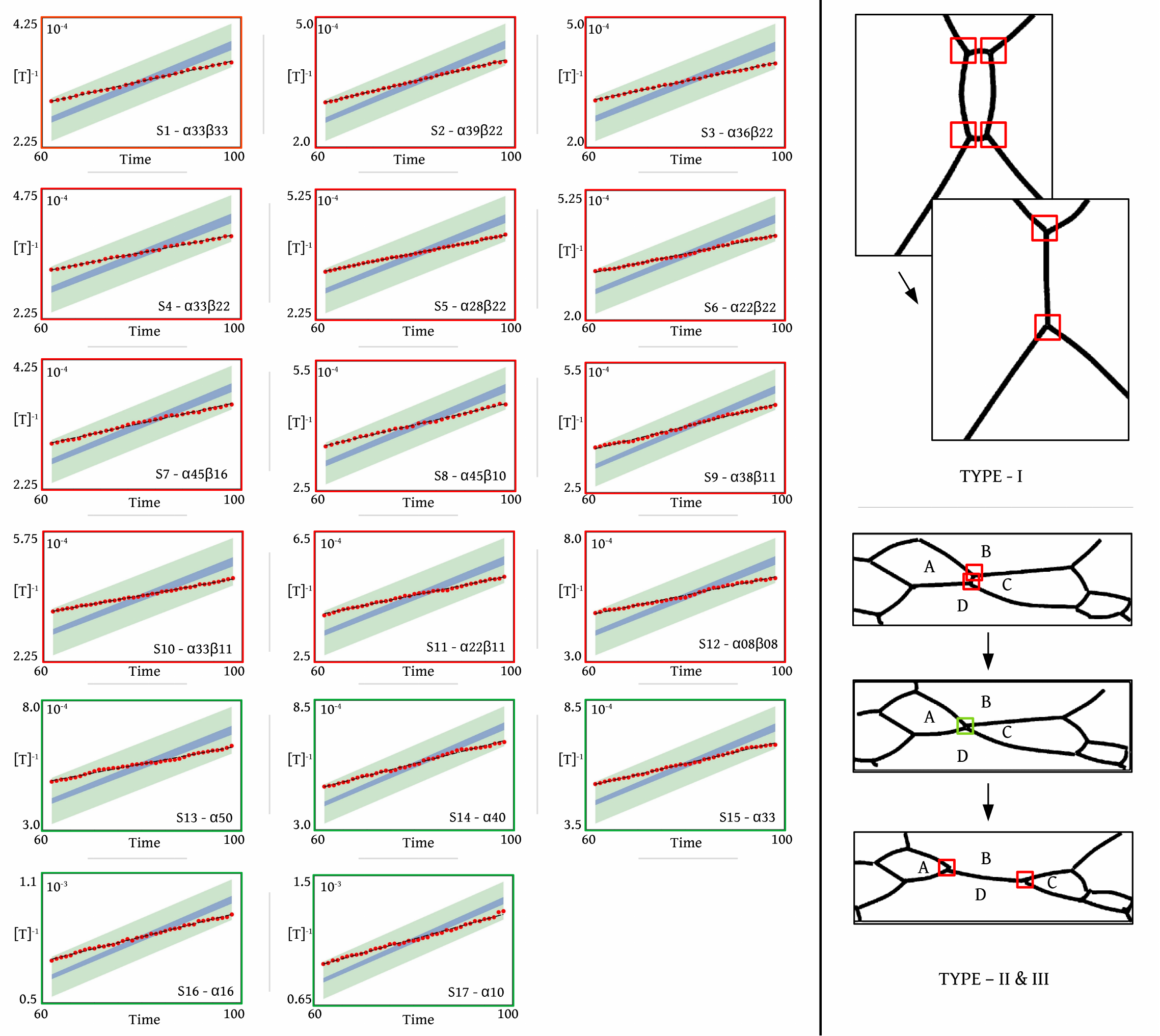}
    \end{tabular}
    \caption{ Reciprocal of the number of triple-junctions, inverse count, [T]$^-1$, is plotted against time for all isotropic duplex and triplex microstructures on the left window. Given the perceivable trend, a linear fit is introduced in all of these plots. The statistical rigor of the inverse count plot is enhanced by appending confidence and prediction intervals, respectively depicted in blue and green. On the right windows, different types of topological events are represented. Evolution of the junctions during these topological events are highlighted. 
    \label{fig:tp2}}
\end{figure}

Grain growth exhibited by 17 multiphase systems - 5 duplex and 12 triplex - is analysed by monitoring the variation in the number of triple junctions in the series of micrographs capturing the associated progressive change. 
Inverse triple-junction count, [T]$^-1$, which is the reciprocal of the number of junction, is ascertained through the junction-detection tool for different multiphase polycrystalline systems. 
The progressive change in the inverse triple-junction count with time (unitless computational timestep~\cite{mittnacht2021morphological}) for all the duplex and triplex systems is presented in Fig.~\ref{fig:tp2}. 
A linear fit, based on the datapoints, is introduced in all these plots.
Disparity between datapoint and the linear fit is rather marginal as discussed in Appendix~\ref{app2:rmse} using root-mean-square-error (RMSE).
In order to enhance the statistical potency of inverse count vs time plot, confidence and prediction intervals are overlaid. 
These intervals are constructed by appending additional datapoints around the existing ones.
In other words, additional datapoints, assumed to be rendered by microstructures with similar phase-fraction but different initial configuration, are considered around the original points. 
These assumed points are normally distributed with the mean and standard deviation dictated by the original data. 
The confidence interval is the mean of the assumed datapoints surrounding the observed ones, whereas the prediction interval encapsulates the additional datapoints themselves. 
The trend exhibited by the confidence and prediction interval indicates linear relation between inverse count and time would persist for all microstructures, even upon repeating the simulations with different initial configurations. 
Put simply, both confidence and prediction interval further substantiate the linear change in the inverse triple-junction count with time. 

Based on the prevailing trend in the temporal change of inverse triple-junction count, across the different systems in Fig.~\ref{fig:tp2}, the grain-growth kinetics can be expressed as 
\begin{equation}\label{key}
[T]^{-1}(t) - [T]_0^{-1}  = k_T t. 
\end{equation}
In above relation, reciprocal of number of triple-junctions at the initial state, and a given instant t, is denoted by $[T]_0^{-1}$ and $[T]^{-1}(t)$, respectively. 
The characteristic kinetic coefficient, $k_T$ in Eqn, quantifies the rate of grain growth in various polycrystalline system in the present triple-junction based approach. 
This coefficient is the slope of linear fit introduced in the respective plots of Fig.~\ref{fig:tp2} to highlight the apparent relation between the number of triple junctions and time during grain growth. 

\subsection{Triple-junctions count and grain disappearance}

Besides the apparent increase in the size of the grains, evolution of polycrystalline during grain growth is accompanied by topological events. 
Stated otherwise, grain growth is characterised not only by the change in the size of the grains but also their number of sides. 
The changes altering the number of sides, or the face class, of the grains are referred as topological events. 
In polycrystalline microstructures, exhibiting grain growth, three types of topological events are realised. 
These events are illustrated on the right panel of Fig.~\ref{fig:tp2}. 
Topological events type-II and -III indicate sequence of evolution wherein two grains, A and C, lose a side. 
This act leads to the formation of a quadruple junction, which is energetically-unstable in isotropic systems. 
Consequently, the quadruple dissociates into two triple junctions, enabling grains B and D increase its number of sides. 
In other words, type-II and III events essentially capture the loss of a side by two abutting grains, which is eventually gained by their neighbours.
Type-I event, on the other hand, represents the topological changes associated with the disappearance of a grain. 
This event, as illustrated in Fig.~\ref{fig:tp2}, reduces the number of sides of top and bottom grains.
When viewed topologically, triple junctions are the sites where the sides of the grains meet. 
Therefore, any change in the topological features of the grains ultimately effects the number of triple junctions. 
Considering type-II and -III topological events, it is event from Fig.~\ref{fig:tp2}, that despite the change in the number of sides to the grains, the total number of triple junctions remain conserved. 
On the other hand, the number of triple junctions reduces from four to two in topological event-I.  
Put simply, topological events are associated either with change in the number of sides or disappearance of grains. 
Events establishing change in the number of sides largely preserve the count of triple junctions. However, topological events accompanying disappearance of grains bring about a noticeable change in number of triple junctions. 

Though grain growth is conventionally perceived as a progressive increase in the average grain-size, this change becomes noticeable only with the disappearance of grains. 
Stated otherwise, the characterising increase in the average grain-size is the reflection of incremental disappearance of grains. 
Since the disappearance of grains, which principally quantifies grain growth, is accompanied by change in the number of triple junctions, the inverse triple-junction count, estimated in the present work (Fig.~\ref{fig:tp2}), offers a reliable measure of the growth kinetics. 

\subsection{In view of conventional estimation}

\begin{figure}
    \centering
      \begin{tabular}{@{}c@{}}
      \includegraphics[width=1.0\textwidth]{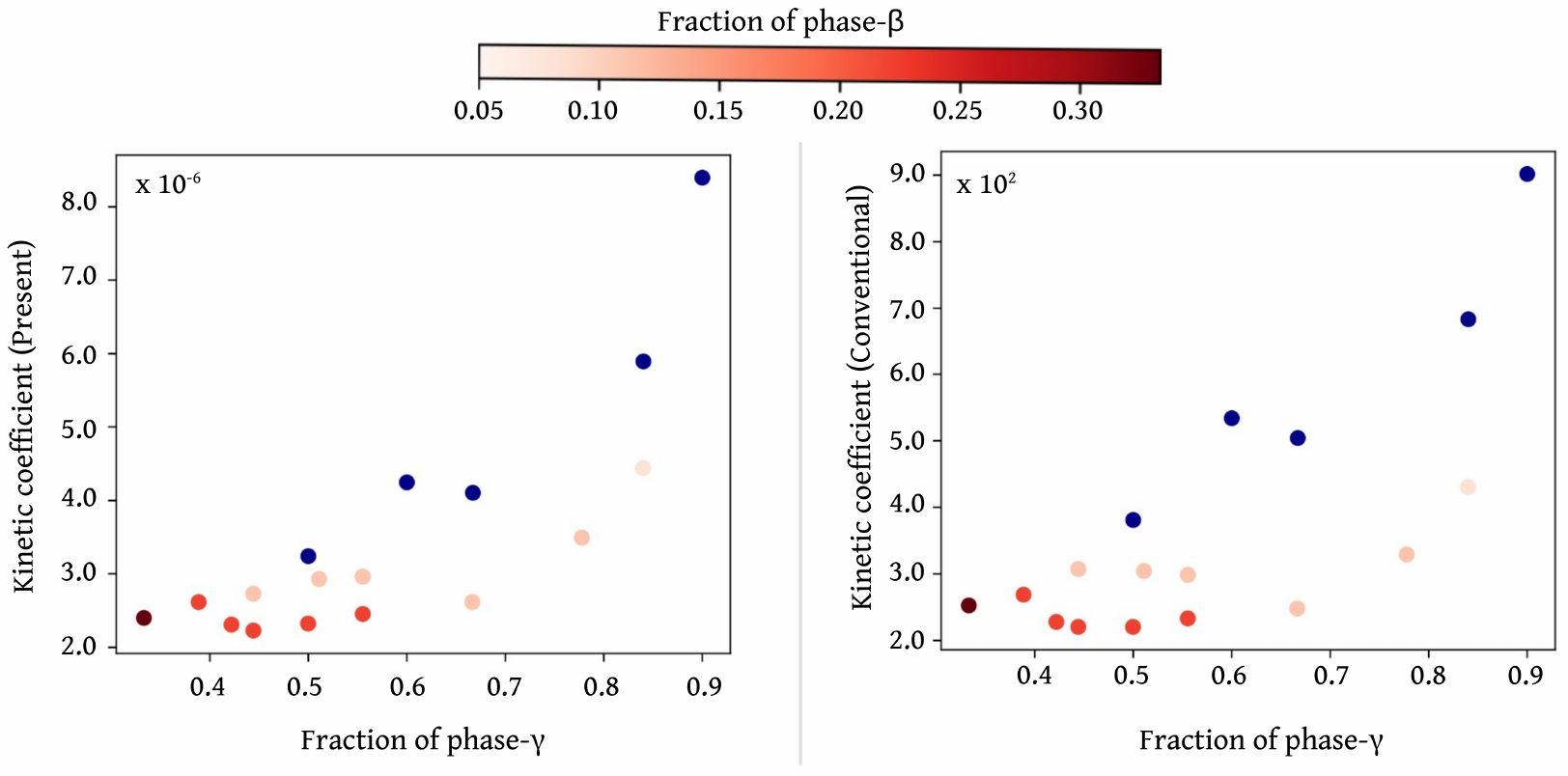}
    \end{tabular}
    \caption{ Kinetic coefficients ascertained from the slope of the inverse count vs time plot in Fig.~\ref{fig:tp2} is presented across corresponding coefficients calculated from the temporal change in average radius. These plots include kinetic coefficients of all isotropic duplex- and triplex-microstructures. 
    \label{fig:conparison}}
\end{figure}

The characteristic kinetic-coefficient of the present triple-junction based approach, estimated as the slope of the linear fit reflecting the temporal change in inverse count $[T]^{-1}(t)$, is shown in Fig.~\ref{fig:conparison}.
This illustration essentially encompasses grain-growth kinetics of all isotropic polycrystalline microstructures. 
Beside this representation, to its right, the kinetic coefficients estimated through the conventional approach is included in Fig.~\ref{fig:conparison}. 
\textcolor{black}{This conventional kinetic-coefficient $(k_R)$ is estimated from the relation, $<R(t)>^n - <R_0>^n = k_R t$, where the kinetic exponent, $n=3$, and the average radius of the polycrystalline system at the initial ($t=0$) and at a given state ($t$) is respectively denoted by  $<R_0>$ and $<R(t)>$~\cite{amos2022high}. 
Here it is vital to note that, while $n=2$ generally reflects the rate of grain growth in isotropic polycrystalline microstructures, in the present systems wherein the evolution is dictated by long-range diffusion, the exponent ($n$) assumes the value of 3~\cite{amos2020multiphase}. }
Given the grain growth in these various systems have been numerically modelled, the average radius is estimated from the phase-field based pixel data~\cite{perumal2017phase}.
Both representation of kinetic coefficients in Fig.~\ref{fig:conparison}, irrespective of the underlying approach, informs the relative growth-rate exhibited by a given polycrystalline microstructure. 
The noticeable similarity in the positions of the multiphase polycrystalline microstructures across the two plot in Fig.~\ref{fig:conparison} indicates the overall agreement between the conventional and triple-junction based estimation of growth rate. 
It is vital to note that, given the fundamental difference in estimating the kinetics and the factors associated with it, an identical spacing between the points cannot be expect across the two plots. However, similar distribution of the points, in Fig.~\ref{fig:conparison}, explicates the ability of the triple-junction based approach to render a meaningful description of the kinetics analogous to the conventional techniques. 

\begin{figure}
    \centering
      \begin{tabular}{@{}c@{}}
      \includegraphics[width=1.0\textwidth]{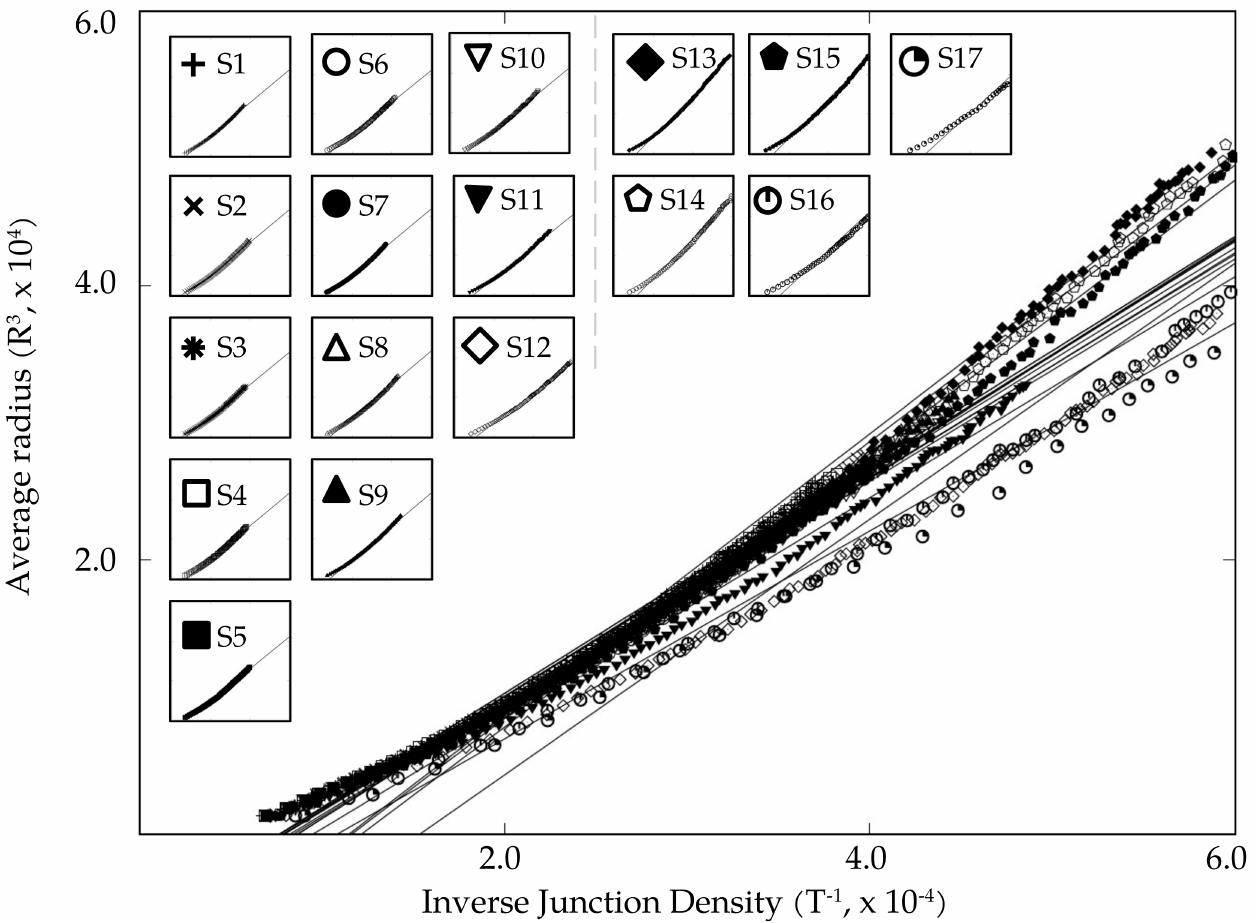}
    \end{tabular}
    \caption{ \textcolor{black}{Progressive change in the inverse triple-junction count ([T]$^{-1}$) is plotted against the corresponding evolution of the average radius ($<R>^{3}$). Each set of points pertaining to a multiphase microstructure is fitted through a line indicative of the characteristic relation between [T]$^{-1}$ and  $<R>^{3}$. In the subplots, the change in radius with triple-junction counts, along with a linear fit, is individually presented  for all multiphase polycrystalline microstructures. }
    \label{fig:TRcRelation}}
\end{figure}

Though the distribution of kinetic coefficients of different systems across the plots are largely similar in Fig.~\ref{fig:conparison}, a noticeable disparity is observed in one point. 
This \textit{outlying} point pertains to a triplex microstructure, s12 in Fig.~\ref{fig:conparison}, with predominant matrix-fraction (0.84) and scarce minor phases (0.8$\alpha$ and 0.8$\beta$).
Owing to the significant difference in the volume fraction of the phases, grains associated with the minor phases assume distinct positions and configuration similar to pinning-precipitates rather than constituent grains. 
Consequently, the topological evolution of the grains are seemingly affected, which in-turn introduces a deviation in the triple-junction based estimation of growth kinetics. 
In other words, though triple-junction based estimation of growth kinetics can be extended to wide range of systems, its outcome in multiphase microstructure with exceedingly low fraction of minor phases is rather not accurate.

\subsection{\textcolor{black}{Junctions and grain size}}

\begin{figure}
    \centering
      \begin{tabular}{@{}c@{}}
      \includegraphics[width=1.0\textwidth]{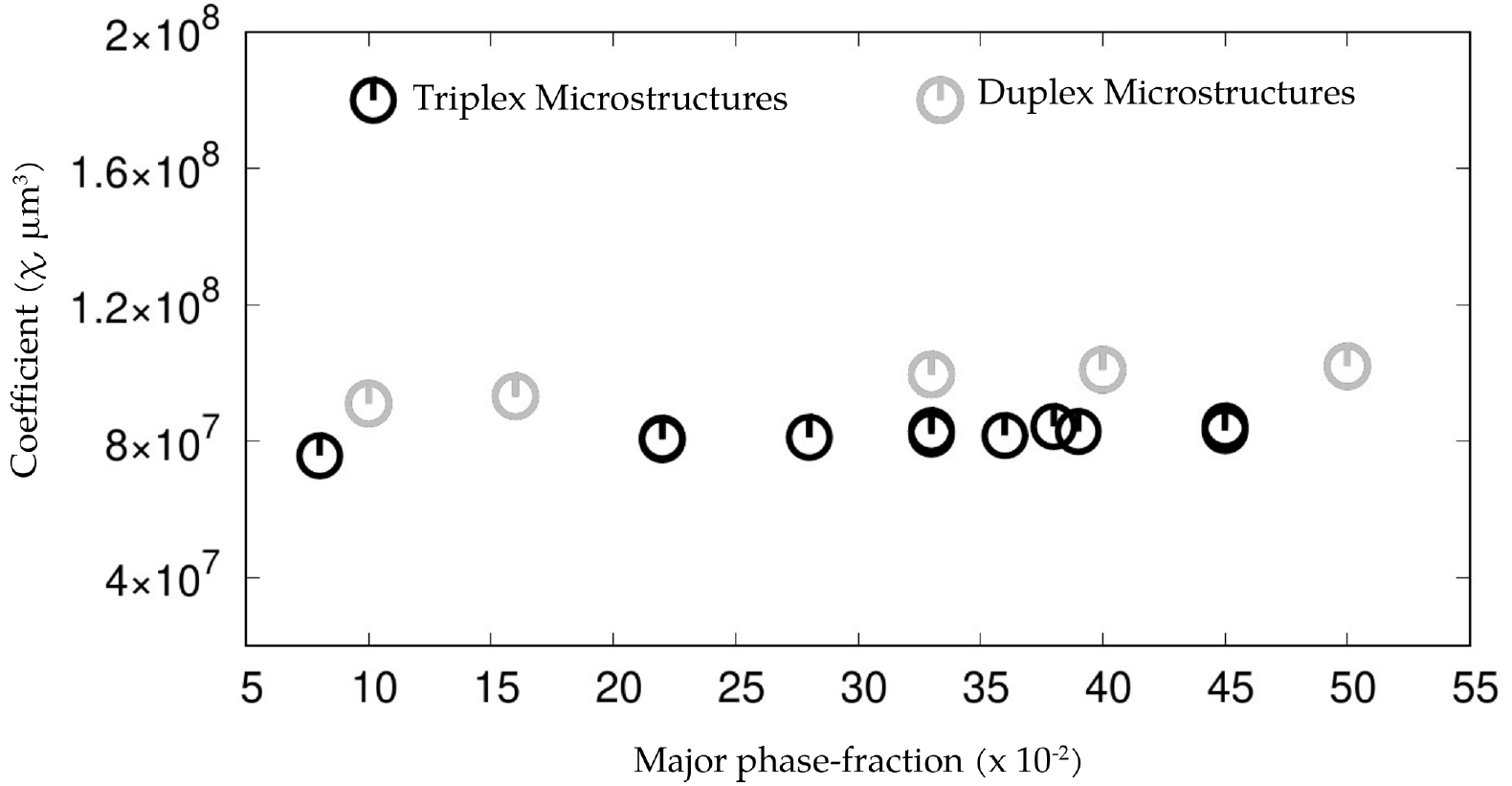}
    \end{tabular}
    \caption{ \textcolor{black}{Coefficient relating the inverse triple-junction counts ([T]$^{-1}$) to the average radius ($<R>^{3}$), estimated from the corresponding linear fit, for various duplex and triplex microstructures. }
    \label{fig:slopeTcR}}
\end{figure}

\textcolor{black}{An exclusive focus on junctions to estimate the growth kinetics means the present approach rather fails to offer reasonable insight on the average grain size of the polycrystalline microstructure. 
This limitation apparently reflecting a rigid framework wherein the outcomes of the current technique cannot be converted (or related) to the kinetics ascertained from the conventional average-radius based approaches. 
In order to reduce the rigidity of the present treatment, and enhance its versatility, the change in the inverse junction count is analysed along with the progressive increase in average radius ($<R>^{3}$).
Stated otherwise, the progressive evolution of the average radius ($<R>^{3}$) and triple-junction counts ([T]$^{-1}$), accompanying the grain growth, is cumulatively studied and presented in Fig.~\ref{fig:TRcRelation}.
This investigation unravels a linear relation between the junction count ([T]$^{-1}$) and average radius ($<R>^{3}$) for all microstructures, irrespective of the number of constituent phases and their fractions, which can be expressed as 
\begin{equation}\label{key2}
 <R(t)>^{3} = \chi [T]^{-1}(t) + \text{A}([T]_0^{-1}).
\end{equation}
The coefficient $\chi$, in the above expression, relates the temporally varying radius to the junction count.  
This parameter ($\chi$) is ascertained from the linear fit of the junction count ([T]$^{-1}$) and average radius ($<R>^{3}$) for all multiphase polycrystalline microstructures and is presented in Fig.~\ref{fig:slopeTcR}.
Though the linear trend is independent of the phase fraction, the coefficient assumes noticeably different values across the duplex and triplex microstructures. 
While in triplex microstructures the coefficient is around $8.5\times10^7 \mu m^{3}$, it increases to $\chi= 9.7\times10^7 \mu m^{3}$ in two-phase systems.
This disparity in the value of the relating coefficient can be attributed to the significantly different growth kinetics exhibited by the duplex and triplex microstructures. 
Besides enabling the conversion of the junction-based outcomes, the cumulative analysis of the triple junctions and the average radius accounts for the huge difference (order of $10^8$) in the kinetic coefficient of the present and the conventional approach.}

\subsection{\textcolor{black}{Bounds of the junction-based treatment}}\label{bounds}

\textcolor{black}{The present junction-based approach of ascertaining kinetics offers an alternate route for both considering and correspondingly, quantifying the evolution of polycrystalline microstructures during grain growth. 
This approach can directly be adopted to gain a comprehensive understanding on the relative differences in the grain-growth rate of a various microstructures, as shown in Fig.~\ref{fig:conparison}. 
Furthermore, by extending the current measuring technique, real-time change in the number of triple (or quadruple) junctions during in-situ studies can be accurately determined.
Such extension, besides obviating the need for any dedicated apparatus, would track the growth-kinetics all through the evolution across the entire sample.
The exclusive focus on the junctions, despite its benefits, rather implicitly precludes the present approach from offering insights on a key parameter associated with conventional treatments. 
In other words, insights on the rate of grain growth conventionally stem from the area (size) of the grains, and subsequently, unravel the distribution of the size in a polycrystalline system.
Present approach, reflecting its principal consideration of junctions, conveys no real information on the grain size distribution.
}

\textcolor{black}{By relating the growth kinetics to a quantity that is restricted to the two-dimensional representation of the polycrystalline system, the present approach limits itself to the analysis of 2D microstructures. 
Stated otherwise, the kinetics of grain growth in the current approach is ascertained by detecting and counting triple junctions which appear as a zero-dimensional object in the two-dimensional microstructures. 
On the other hand, in three-dimension, these features are one-dimensional entities wherein the grain-boundary surfaces meet. 
Consequently, the present junction-based approach cannot be directly extended to three-dimensional microstructures to determine growth kinetics. 
However, when the intent is to understand the relative differences in the growth kinetics across various microstructures, the current technique offers accurate results more efficiently from the two-dimensional micrographs (Fig.~\ref{fig:conparison}).
}

\section{Conclusion}

Estimating grain-growth kinetics, which captures the rate of change in the size of the constituent grains, is pivotal in polycrystalline systems. 
This importance is primarily because the behaviour of polycrystalline materials is noticeably influenced by the average size of the grains. 
Techniques generally adopted to estimate grain-growth kinetics involves measuring average grain-sizes of the micrographs reflecting temporal evolution of the polycrystalline systems. 
Such techniques are straightforward, easy to comprehend, and offer reliable outcome. 
On the other hand, besides being rather arduous, the conventional approaches generally offer a localised estimation of the size and consequently, demand several iterations. 
Even though the various steps of the conventional techniques can be (and are) automated, the fundamental limitations emerging from the localised focus continue to exist. 
Despite the increasing role played by artificial intelligence in deciphering intricate microstructures, these advancements are yet to shift the paradigm of grain-growth measurement. 
In the present work, an approach based on object-detection is shown to offer an alternate for estimating growth kinetics in polycrystalline systems.

The progressive increase in the average grain-size of the polycrystalline material during grain growth is primarily due to incremental disappearance of grains. 
In other words, tracking the gradual decrease in the number of grains would render a convincing estimation of the growth kinetics.
However, such tracking would involve monitoring the evolution of numerous grains of varying morphology simultaneously. 
The change in the number of grains, besides altering the average grain-size, change the microstructural features, particularly the triple junctions. 
Put simply, the disappearance of the grains, which dictate the increase in the average grain-size, thereby dictating growth kinetics, is accompanied by decrease in triple junctions. 
Correspondingly, an object-detection algorithm is extended to realise and count the temporal change in the number of triple junction to render the growth kinetics. 
As opposed to the conventional technique, this triple-junction based approach considers a polycrystalline microstructure holistically rather than locally. 
Moreover, inadequate etching which poses a serious impediment to application of conventional techniques, can be handled by the triple-junction based approach. 
In other words, by offering a growth rate based on the temporal change in the number of distinguishable triple-junctions, the effect of the inadequately etched features are minimised in the present approach. 
The proven ability of the underlying algorithm to operate in real-time suggests the potential applicability of the current technique for the in-situ measurement of grain-growth kinetics~\cite{wu2021real}. 

\textcolor{black}{Ascertaining the growth kinetics from the temporal change in the number density of junctions lends itself to a unique framework in which the present approach operates. 
This framework, as discussed in Sec.\ref{bounds}, seemingly overlooks certain features, particularly grain size (area) and distribution, that constitute the foundation of conventional techniques.
The lack of explicit inclusion of  grain size and distribution in the framework of junction-based approach precludes a direct conversion growth rate determined by one technique to another.
In order to address this gap, and enhance the versatility of the present treatment, attempts will be made in the subsequent works to first establish a relation between the junction count and size of the individual grains. 
A deeper understanding of this relation between the topological and geometrical feature would facilitate in the cross-verification of the growth kinetics estimated from different techniques.
}

%
%

\section*{Appendices}

\subsection*{Appendix 1 - Multicomponent multiphase-field model}\label{App:1}

In multiphase-field modelling of grain growth, each grain is distinguished by a scalar variable called phase-field such the evolution is realised by the spatio-temporal change in this variable not by the interfaces. 
Since the multiphase system is involved in this work, the conventional description of polycrystalline microstructure is extended and distinct phases are associated with the grains. 
Correspondingly, the phase-field variables representing the grains in a multiphase set-up adopt a form
\begin{align}\label{eq:mp2}
\vphi=\Big\{ \underbrace{ \{ \phia^{1},\phia^{2}\dots\phia^{q_{\alpha}} \}}_{\vv{\phia}}, \underbrace{ \{ \phib^{1},\phib^{2}\dots\phib^{q_{\beta}} \}}_{\vv{\phib}}\dots \underbrace{ \{ \phi_{N}^{1},\phi_{N}^{2}\dots\phi_{N}^{q_{N}} \}}_{\vv{\phi_{N}}} \Big\}.
\end{align}
Associating phase to a grain involves assigning a characteristic composition, $\{\vphi(\vc)=\vphi_\alpha(\vc) | \vc=\vv{c}_\alpha\}$, where $\vv{c}_\alpha$ includes all alloying elements. The concentration of the phases are specifically realised and assigned to ensure that chemical equilibrium exists between the different phases and any evolution is primarily governed by the inherent difference in the curvature, \textit{id est} the grain growth.
Concentrations establishing equilibrium across the phases are tabulated in Table.~\ref{tab:mat1}.

Despite the assuming same concentration, the grains of similar phases are distinguished by the characteristic phase-field. These distinct phase-fields avoid coalescence and ensure a stable grain boundary between the chemically-similar grains. 

Based on the characteristic phase-fields establishing multiphase polycrystalline microstructure, the overall energy-density of the system is written as
\begin{align}\label{eq:functional1}
\cal{F}(\vphi,\n \vphi,\vc) & =\cal{F}_{\text{int}}(\vphi,\n \vphi)+\cal{F}_{\text{bulk}}(\vphi,\vc) \\ \nonumber
&=\int_{V}f_{\text{int}}(\vphi,\n \vphi)+f_{\text{bulk}}(\vphi,\vc)\diff V,
\end{align}
where $\int_{V}f_{\text{int}}(\vphi,\n \vphi)$ and $f_{\text{bulk}}(\vphi,\vc)\diff V$ represent the energy-density contribution of the diffuse interface and bulk phase-associated grains, respectively. 
The spatio-temporal evolution of phase-field which translates to the evolution of the grains is modelled by minimising the overall energy-density of the system. 
Therefore, phase-field evolution is solved by 
\begin{equation}
\tau\epsilon\frac{\partial \phi_{\alpha}^{m}}{\partial t}  = -\frac{\partial \cal{F}(\vphi,\n \vphi,\vc)}{\partial \phi_{\alpha}^{m}} =\epsilon\left[ \n \cdot \frac{\partial a(\vphi, \n \vphi)}{\partial \n \phi_{\alpha}^{m}} - \frac{\partial a(\vphi, \n \vphi)}{\partial \phi_{\alpha}^{m}} \right] - \frac{1}{\epsilon} \left[ \frac{\partial w(\vphi)}{\partial \phia^{m}} \right] - \left[ \frac{{\partial}f^{\alpha}_{m}(\vc^{\alpha},\phia^{m})}{\partial \phia^{m}} \right] - \Lambda,
 \label{eq:ph_evo1}
\end{equation}
where $\phi_{\alpha}^{m}$ indicates grain-$m$ associated with phase-$\alpha$. Terms $w(\vphi)$ and $a(\vphi, \n \vphi)$ denote penalising potential and gradient energy with $f^{\alpha}_{m}(\vc^{\alpha},\phia^{m})$ indicating equilibrium free-energy of grain-$m$. At any given point, sum of all the constituent phase-fields should be one. This characteristic criterion of the multiphase-field approach is ensured by the Lagrange multiplier, $\Lambda$.

\begin{table*}
  \caption{Concentration of phases establishing chemically equilibrium system} \label{tab:mat1} 
 \centering
 \begin{tabular}{c | c | c }
 
  Phase 		& Independent component-$i$ 	& Independent component-$j$ \\ [0.5ex]
  \hline
  (Binary)		&				&   \\
  Phase-$\alpha$	& 0.1				& - \\
  Phase-$\beta$		& 0.9				& - \\
  \hline
  (Ternary)		&				&   \\
  Phase-$\alpha$	& 0.05				& 0.05 \\
  Phase-$\beta$		& 0.05				& 0.9 \\
  Phase-$\gamma$	& 0.9				& 0.05\\
 \end{tabular}
\end{table*} 

The spatio-temporal evolution of the dynamic variable, chemical potential ($\mu_{i}$), establishing equilibrium and curvature-based migration of the chemical components reads
\begin{equation}
\frac{\partial \mu_{i}}{\partial t}=\left \{ \n\cdot\left[ \sum_{j=1}^{K-1} \vv{M}(\vphi) \n \mu_{j}\right ] - \sum_{\alpha}^{N} \sum_{m}^{q} c_{i}^{\alpha}\frac{\partial \phia^{m}}{\partial t} \right \} \left [ \sum_{\alpha}^{N} \sum_{m}^{q} h(\phia^{m}) \frac{\partial c_{m:i}^{\alpha}}{\partial \mu_{j}} \right ]_{ij}^{-1},
 \label{chempot_ev}
\end{equation}
where $\vv{M}(\vphi)$ is matrix encompassing mobility of the migration chemical species.
Table.~\ref{tab:mat2} lists the constants assumed to model all multiphase polycrystalline systems.

\begin{table*}
  \caption{Material parameters adopted to model duplex and triplex microstructures} \label{tab:mat2} 
 \centering
 \begin{tabular}{c | c | c }
  Parameter 		& Symbol 	& Value \\ [0.5ex]
  \hline
  Grain boundary energy	& $\bar{\gamma}_{\alpha \alpha}=\bar{\gamma}_{\beta \beta}\bar{\gamma}_{\gamma \gamma}$				& 1.0 $\text{Jm}^2$ \\
Interphase boundary energy	& $\bar{\gamma}_{\alpha \beta}=\bar{\gamma}_{\beta \gamma}=\bar{\gamma}_{\alpha \gamma}$	&  1.0 $\text{Jm}^2$\\
Bulk diffusivities & $D^{\alpha}=D^{\beta}=D^{\gamma}$ & 1.0 $m^2s^{-1}$\\
  \hline
 \end{tabular}
\end{table*} 

Despite the varying phase-fractions, all multiphase polycrystalline microstructures are intialised and their evolutions are modelled in two-dimensional domains of identical configurations
Through finite-difference scheme, the two-dimensional domains are discretised into $2048 \times 2048$ grids of dimensions, $\Delta$x = $\Delta$y = $5 \times 10^{-7}$m uniformly. 
Polycrystalline setup over these discretised grids are established by Voronoi tessellation rendering approximately $10000$ grains. 

\begin{table*}
  \caption{Parameters associated with the multiphase-field simulations} \label{tab:mat3} 
 \centering
 \begin{tabular}{c | c | c }
  Parameter 		& Symbol 	& Value \\ [0.5ex]
  \hline
  Timestep width (\text{No unit}~\cite{amos2020multiphase})	&  $\Delta$t 				& 1.0  \\
Interface-width pre-factor	& $\epsilon$	&  0.2 $~\mu \text{m}$~\cite{mittnacht2021morphological}\\
Relaxation parameter & $\tau$ & 1.0 ~$\text{Jsm}^{-4}$\\
  \hline
 \end{tabular}
\end{table*}

\begin{figure}
    \centering
      \begin{tabular}{@{}c@{}}
      \includegraphics[width=0.9\textwidth]{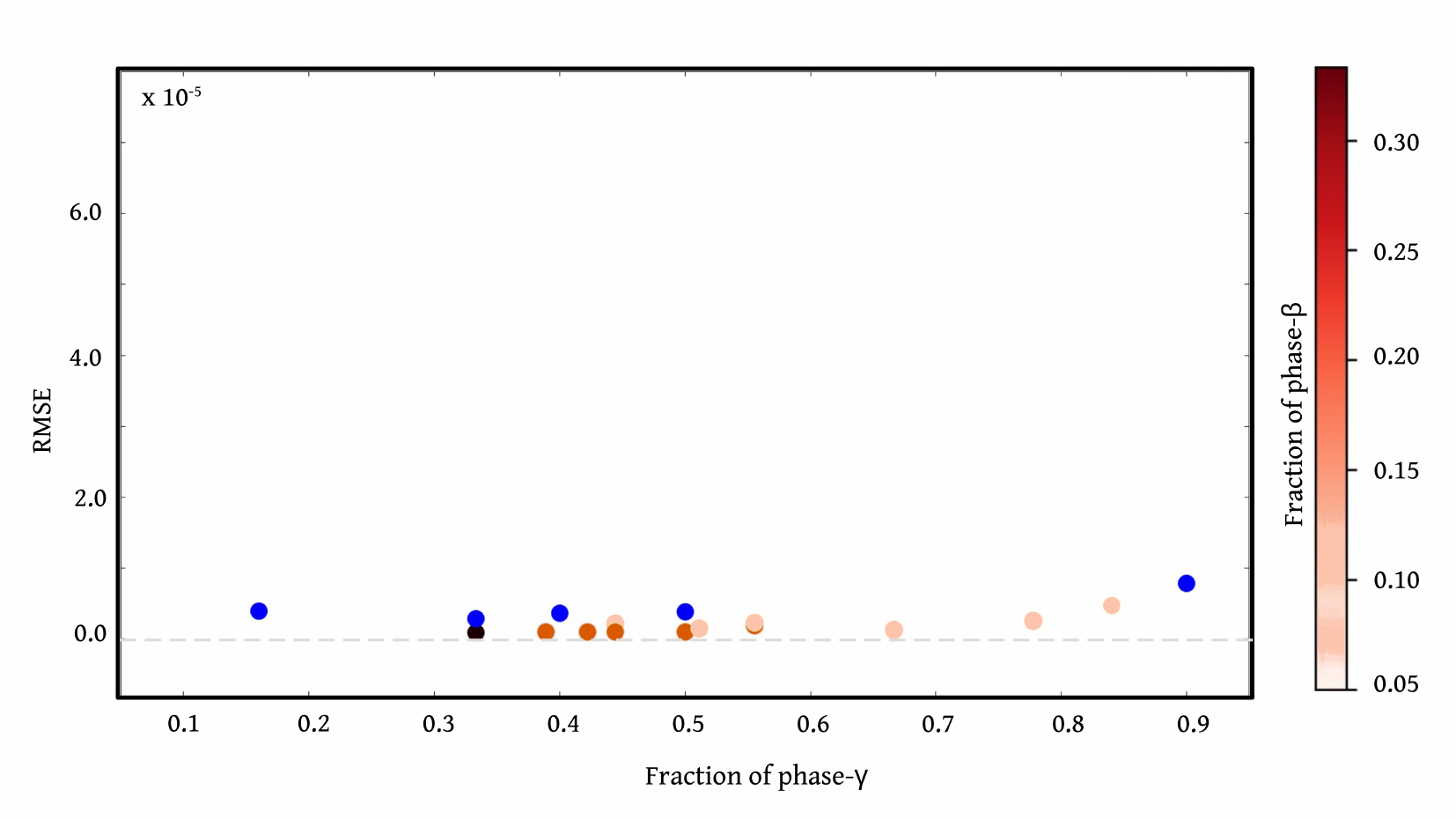}
    \end{tabular}
    \caption{ The deviation of the datapoints from fitting introduced in all inverse count vs time plots presented in Fig.~\ref{fig:tp2} is quantified as RMSE (Root-Mean-Squared-Error). 
    \label{fig:RMSE_TJ}}
\end{figure}

Microstructural evolution of the multiphase polycrystalline system is modelled by solving temporal evolution of the dynamic variables - phase-field and chemical potential of components - by forward-marching Euler's scheme over the homogeneously discretised domain.
The simulation domain is decomposed into sections of smaller dimensions, and handled simultaneously by Message Passing Interface (MPI).
The involvement MPI ensures efficient use of the computational resources. 

\subsection*{Appendix 2 - Linear fitting of inverse triple-junction datapoints}\label{app2:rmse}

Upon of introducing a linear fit to the datapoints of all microstructures in Fig.~\ref{fig:RMSE_TJ}, the disparity between the individual point and the fitted line is estimated as RMSE (Root-Mean-Squared-Error). 
These deviations between fitting and datapoints, quantified as RMSE, for different microstructures are illustrated in Fig.~\ref{fig:RMSE_TJ}. 
This depiction unravels the RMSE value for all the multiphase systems fall within $1.0 \time 10^{-5}$. In other words, irrespective of the phase-fractions, minimal disparity exist between the linear fitting and datapoints in all multiphase systems. 
Consequently, the linear relation is adopted to describe the change in inverse triple-junction count with time.

\section*{\textcolor{black}{Data Availability}}
\textcolor{black}{The source code for detecting junctions in polycrystalline microstructures and sample images employed in the present work are made available at \\
\url{https://github.com/theormets/Junction-based-Grain-Growth-Measurement}}

\section*{Declaration of Interest}
The authors declare that they have no known competing financial interests or personal relationships that could have appeared to influence the work reported in this paper.

\section*{Acknowledgments}

PGK Amos thanks the financial support of the SCIENCE \& ENGINEERING RESEARCH BOARD (SERB) under the project SRG/2021/000092.

\end{document}